# Molecular gates in mesoporous bioactive glasses for the treatment of bone tumors and infection

Lorena Polo,[a,c,‡] Natividad Gómez-Cerezo,[b,c,‡] Elena Aznar,[a,c] José-Luis Vivancos,[a,c] Félix Sancenón,[a,c] Daniel Arcos,[b,c]* María Vallet-Regí,[b,c]* and Ramón Martínez-Máñez,[a,c]*

[a] Instituto Interuniversitario de Investigación de Reconocimiento Molecular y Desarrollo Tecnológico (IDM) Unidad Mixta Universitat de València–Universitat Politècnica de València, Camino de Vera s/n, 46022, Valencia, Spain
Phone: +34 963877343, Fax: +34 963879349.

[b] Departamento de Química Inorgánica y Bioinorgánica, Facultad de Farmacia, Universidad Complutense de Madrid
Plaza Ramón y Cajal s/n 28040, Madrid, Spain
Phone: +34 913941843, Fax: +34 913941786

[c] CIBER de Bioingeniería Biomateriales y Nanomedicina (CIBER-BBN), Spain

[‡] These authors contributed equally to this work.

*Corresponding authors e-mail: rmaez@qim.upv.es, arcosd@ucm.es, vallet@ucm.es

**Abstract**

Silica mesoporous nanomaterials have been proved to have meaningful application in biotechnology and biomedicine. Particularly, mesoporous bioactive glasses are recently gaining importance thanks to their bone regenerative properties. Moreover, the mesoporous nature of these materials makes them suitable for drug delivery applications, opening new lines in the field of bone therapies. In this work, we have developed innovative nanodevices based on the implementation of adenosine triphosphate (ATP) and ε-poly-L-lysine molecular gates using a mesoporous bioglass as an inorganic support. The systems have been previously proved to work properly with a fluorescence probe and subsequently with an antibiotic (levofloxacin) and an antitumoral drug (doxorubicin). The bioactivity of the prepared materials has also been tested, giving promising results. Finally, in vitro cell culture studies have been carried out; demonstrating that this gated devices can provide useful approaches for bone cancer and bone infection treatments.

## 1. Introduction

Mesoporous bioactive glasses (MBGs) are a new generation of bioceramics designed for bone grafting and skeletal regenerative therapies [1,2]. These biomaterials exhibit the bone regenerative properties of bioactive glasses [3], but significantly increased due to their high surface area and porosity [4,5]. In addition, MBGs possess ordered mesoporous structures similar to those exhibited by pure silica mesoporous materials, which makes them excellent candidates as matrixes in drug delivery applications [6–8]. This synergy between osteogenic properties and local drug delivery capabilities is called to play a main role in field of skeletal therapies in near future. Several studies have been carried out to evaluate the behavior of MBGs as drug delivery systems [9], which have demonstrated that MBGs can release drugs following classical diffusion mechanism [10]. However, the real potential of MBGs remains still unknown. The possibility of supplying stimuli-responsive behavior to MBGs, in such a manner that they release the payload only when the pathological process occurs is unexplored.

From a different point of view, and in the context of on-command delivery, mesoporous silica has been used as an effective support for the development of controlled-release nanodevices because of their unique characteristics, such as high homogeneous porosity, inertness, robustness, thermal stability, and high loading capacity [11–14]. Consequently, a number of nanodevices for on-command delivery that can be triggered by target chemical [15–21], physical [22–26] or biochemical stimuli [27–34] have been designed recently. However these gated systems have been mainly developed in individual pure silica particles (usually nanometric) [35], and gated functionalities in bioactive compositions for bone regenerative purposes are practically unknown [36].

In order to explore new bone regeneration strategies, stimuli-responsive 3D macroporous scaffolds have been recently prepared by means of incorporating gated $SiO_2$ mesoporous nanoparticles within a polymeric matrix [37]. However, as far as we are aware, none of these systems have been designed for combining bone regenerative properties and on-command drug delivery. The biological response of MBGs can be partially tailored by controlling the supramolecular mechanisms that rules the synthesis. For instance, we have recently demonstrated that the differentiation of pre-osteoblast toward osteoblast phenotype can be enhanced in contact with MBGs prepared with F68, a $(EO)_{78}$-$(PO)_{30}$-$(EO)_{78}$ triblock copolymer which acts as a structure directing agent [38]. However, the complexity to incorporate the nanogates onto multicomponent $SiO_2$-$CaO$-$P_2O_5$ systems has hindered perhaps the development of stimuli-responsive MBGs up to date. Nonetheless, the design of gated MBGs is highly appealing and might found a number of applications in advanced regenerative therapies.

Based on these concepts, it was in our aim to demonstrate the possibility of preparing stimuli-responsive MBGs, via the implementation of tailor-made gated ensembles on the surface of bioactive glasses. The final goal was to develop bioactive glasses able to regenerate bone tissue in bone defects while treating the causal pathology of such a defect, for instance bone infection or tumor extirpation. Therefore the mesopores of a selected MBG were capped with two different enzyme-responsive molecular gates based in the use of adenosine triphosphate (ATP) and ε-poly-L-lysine as caps (*vide infra*), which allowed controlled cargo release in the presence of alkaline phosphatase (ALP) and proteolytic enzymes, respectively. Proteolytic activity can be observed in the presence of different infectious pathogens as *Escherichia coli* [39,40], whereas high levels of serum ALP have prognostic significance of osteosarcoma scenarios [41]. We envision that such tailor-designed systems may have potential applications for the treatment of bone tissue defects commonly associated to osteomyelitis and bone tumors extirpation.

## 2. Materials and methods

### 2.1 Chemicals

The chemicals tetraethyl orthosilicate (TEOS) (98%), triethyl phosphate (TEP) (99%), calcium nitrate $Ca(NO_3)_2 \cdot 4H_2O$ (99%), F68 $(EO)_{78}$-$(PO)_{30}$-$(EO)_{78}$ triblock copolymer, tris(2,2'-bipyridyl)ruthenium(II) chloride hexahydrate ($[Ru(bpy)_3]^{2+}$), 3-[2-(2-aminoethylamino)ethylamino]propyl-trimethoxysilane (**N3**), adenosine 5'-triphosphate disodium salt hydrate (ATP) (99%), N-(3-Dimethylaminopropyl)-N'-ethylcarbodiimide hydrochloride (EDC), alkaline phosphatase (ALP) from bovine intestinal mucosa(buffered aqueous glycerol solution, ≥6,500 DEA units/mg protein), 3-(triethoxysilyl)propyl isocyanate, acetonitrile anhydrous (99.8%), hydrochloric acid (37%), levofloxacin (98%), Dulbecco's Modified Eagle'sMedium and pronase enzyme from *S. griseus* were purchased from Sigma–Aldrich Química S.A. Triethylamine (TEA) (98%) was purchased from J.T. Baker Chemicals. ε-Poly-L-lysine was purchased from Chengdu Jinkai Biology Engineering Co., Ltd. Fetal Bovine Serum (FBS) was obtained from Gibco, BRL.LB medium was provided from Laboratorios Conda. L-glutamine, penicillin and streptomycin were purchased from BioWhittaker-VWR Europe. Doxorubicin used was obtained from the European Pharmacopoeia Reference Standard (Council of Europe EDQM).

### 2.2 General Techniques

FTIR spectroscopy was carried out with a Nicolet Magma IR 550 spectrometer. TEM images were obtained with a 100 kV Jeol JEM-1400 PLUS microscope. SEM images and EDS results were obtained with a 20kV JEOL F-6335 microscope. Differential thermal analysis was done in a TG/DTA Seiko SSC/5200 thermobalance between 50 °C and 1000 °C at a heating rate of 10°C·min$^{-1}$. The textural properties of the calcined materials were determined by nitrogen adsorption porosimetry by using a Micromeritics ASAP 2020 porosimeter. To perform the $N_2$ adsorption measurements, the samples were previously degassed under vacuum for 15 hours, at 150 °C. The surface area was determined using the Brunauer-Emmett-Teller (BET) method. The pore size distribution between 0.5 and 40 nm was determined from the adsorption branch of the isotherm by means of the Barret-Joyner-Halenda (BJH) method. $^1H \rightarrow {}^{29}Si$ and $^1H \rightarrow {}^{31}P$ CP (cross-polarization)/MAS (magic-angle-spinning) and single-pulse (SP) solid-state nuclear magnetic resonance (NMR) measurements were performed to evaluate the different silicon and phosphorus environments in the synthesized samples. The NMR spectra were recorded on a Bruker Model Avance 400 spectrometer. Samples were spun at 10 kHz for $^{29}Si$ and 6 kHz in the case of $^{31}P$. Spectrometer frequencies were set to 79.49 and 161.97MHz for $^{29}Si$ and $^{31}P$, respectively. Chemical shift values were referenced to tetramethylsilane (TMS) and $H_3PO_4$ for $^{29}Si$ and $^{31}P$, respectively. The CP spectra were obtained using a proton enhanced CP method, at a contact time of 1 millisecond. The time period between successive accumulations was 5and 4seconds for $^{29}Si$ and $^{31}P$, respectively, and the number of scans was 10000 for all spectra. Solid-state $^{13}C$ spectra were obtained with a Bruker Model Avance 400 spectrometer 75.46 MHz. For the cell proliferation test, the absorbance was measured using a Helios Zeta UV-VIS spectrophotometer.

**2.3 Synthesis of mesoporous bioactive glasses functionalized**

**2.3.1 Synthesis of the mesoporous bioactive glass (S1)**

85%$SiO_2$-10%CaO-5%$P_2O_5$ (% mol) mesoporous glass (**S1**) was synthetized by evaporation induced self-assembly (EISA) method, using F68 $(EO)_{78}$-$(PO)_{30}$-$(EO)_{78}$ triblock copolymer as structure directing agent. TEOS, TEP and calcium nitrate $Ca(NO_3)_2·4H_2O$ were used as $SiO_2$, $P_2O_5$ and CaO sources, respectively. In a typical synthesis, 2 g of F68 were dissolved in 30 g of ethanol with 0.5 ml of HCl 0.5 M solution at room temperature. Afterward, 3.70 g of TEOS, 0.34 g of TEP and 0.49 g of $Ca(NO_3)_2·4H_2O$ were added under stirring in 3 hours intervals. The resulting solution was stirred during 12 hours and casted into Petri dishes (9 cm in diameter). The colorless solution was evaporated at 30 °C during 11 days. Eventually, the dried gels were removed as homogeneous and transparent membranes, and heated at 700 °C for 3 hours under air atmosphere. Finally the MBG powder was gently milled and sieved, collecting the particle size fraction below 40 µm.

### 2.3.2 Synthesis of S2

1 g of **S1** was suspended in 30 mL of anhydrous acetonitrile under inert atmosphere. Then, 1 mL (3.88 mmol) of 3-[2-(2-aminoethylamino)ethylamino]propyl-trimethoxysilane (**N3**) was added and the mixture was stirred for 5.5 hours. Finally, the solid was filtered, washed with $H_2O$ repeatedly and dried under vacuum for 12 hours.

### 2.3.3 Synthesis of S3

200 mg of solid **S2** were suspended in a solution of EDC 0.1 M and ATP 1.8 M, previously adjusted to pH 7 with NaOH. The suspension was stirred for 6 hours at room temperature. The resulting solid was filtered and dried under vacuum for 12 hours.

### 2.3.4. Synthesis of S4

In a typical synthesis, 1 g of **S1** was suspended in 50 mL of methanol, inside a round-bottom flask under inert atmosphere. Then, an excess of 3-(triethoxysilyl)propylisocyanate (isocyanate) (1 mL, 4 mmol) was added and the final mixture was stirred for 5.5 hours at room temperature. The solid was filtered, washed with H$_2$O repeatedly, and dried under vacuum for 12 hours.

### 2.3.5. Synthesis of S5

A solution of 1 g of ε-poly-L-lysine (0.2 mmol) and TEA (1.6 mL, 11.5 mmol) in 20 mL of methanol was added onto the solid **S4**, and the mixture was stirred for 2 hours. Finally, the solid was filtered off and dried under vacuum.

### 2.3.6. Synthesis of S3-Ru and S5-Ru

In order to carry out the proof of concept about the functioning of ATP and ε-poly-l-lysine based gates, both systems were loaded with a fluorescence dye. For this purpose 2 g of **S1** were suspended in a solution of 1.6 g tris(2,2'-bipyridil)ruthenium(II) chloride hexahydrate ([Ru(bpy)$_3$]$^{2+}$ dye in 70 mL of anhydrous acetonitrile in a round-bottomed flask. Then, 10 mL of acetonitrile were distilled with a dean-stark, in order to remove the possible water present in the pores of the solid. Afterwards, the mixture was stirred at room temperature during 24 hours, in order to achieve maximum loading in the pores of the MBG scaffolding. The resulting solid was then filtered and dried under vacuum for 12 hours. Subsequently, the solid was functionalized and capped as **S3** and **S5** respectively, as described above. Once dried, the solids were suspended in 5 mL of water and stirred in short washing, in order to remove the excess of dye remaining on the surface of the materials. The solids were filtered again and dried under vacuum for 12 hours.

### 2.3.7. Synthesis of S3-Levo and S5-Levo

In order to test the functioning of ATP and ε-poly-L-lysine based gates as on command antibiotic delivery devices, both systems were loaded with levofloxacin. For this purpose 800 mg of **S1** were suspended in a solution of 400 mg (1.11 mmol) of levofloxacin in 20 mL of methanol inside a round-bottom flask under inert atmosphere.

The mixture was stirred at room temperature during 24 hours to achieve maximum loading in the pores of the MBG. Subsequently, the solution was split in two and each part was functionalized in a similar way to **S3** and **S5** respectively. Finally **S3-Levo** and **S5-Levo** were washed in water to remove the excess of antibiotic remaining on the surface.

**2.3.8. Synthesis of S3-Dox**

In order to test the functioning of the ATP based gate as on command antitumoral delivery devices, 30 mg of solid **S2** and 6 mg of doxorubicin were suspended in 5mL of methanol and stirred for 50 hours. Then, the solid was centrifuged and dried under vacuum for 1 hour. Finally the solid was introduced in a solution of EDC 0.1 M and ATP 1.8 M previously adjusted to pH 7 with NaOH. The suspension was stirred for 6 hours at room temperature. The resulting solid was centrifuged, washed several times with water and methanol and dried under vacuum for 12 hours. **Table 1** contains a summary of the synthesized solids.

**2.4 Stimuli-responsive studies**

**2.4.1 Stimuli-responsive studies with S3-Ru and S5-Ru**

4 mg of the **S3-Ru** and **S5-Ru** were suspended in 10 mL of water, and the pH was adjusted to 7.6 with NaOH. Both suspensions were split in two in order to study the dye release in the absence or presence of the corresponding stimuli, i.e. ALP and pronase for **S3-Ru** and **S5-Ru**, respectively. The two samples were stirred at 400 rpm and 37°C, and then several 250µL aliquots were taken for each sample at different times. These aliquots were filtered with PFTE filters (0.22 µm) to monitor the $[Ru(bpy)_3]^{2+}$ release ($\lambda_{ex}$ 454 nm, $\lambda_{em}$ 593 nm) by fluorescence spectroscopy

**2.4.2 Stimuli-responsive studies with S5-Levo**

4 mg of **S5-Levo** were suspended in 10 mL of water, and the pH was adjusted to 7.6 with NaOH. The suspension was split in two in order to study the levofloxacin release in the absence and the presence of *E.coli* (final concentration $10^5$ cells mL$^{-1}$). The two samples were stirred at 400 rpm and 37 °C, and then several 250µL aliquots were taken for each sample at different times. These aliquots were filtered with PFTE filters (0.22 µm) to monitor the levofloxacin release ($\lambda_{ex}$ 292 nm, $\lambda_{em}$ 494 nm) by fluorescence spectroscopy.

### 2.4.3 Stimuli-responsive studies with S3-Dox

4.6 mg of **S3-Dox** were suspended in 2 mL of H$_2$O pH 7.6. Then, 0.5 mL of this suspension was placed on a Transwell permeable support with polycarbonate membrane (0.4 μm). The well was filled with 1.5 mL of H$_2$O pH 7.6 and the suspension was stirred at 37 °C and 100 rpm during all the experiment. To study the enzymatic responsiveness of the solid, the suspension was in contact with ALP on a Transwell. The amount of cargo released was determined by fluorescence spectrometry $\lambda_{exc}$ 490, $\lambda_{em}$ 514 nm, and the solution outside the Transwell insert was replaced with fresh medium, with or without ALP.

### 2.5 Bioactivity assays with solids S3 and S5

Assessments of in vitro bioactivity were carried out on **S3** and **S5** solids. For this purpose, 40 mg of the solids were soaked into 6 mL of filtered simulated body fluid (SBF) [46] in polyethylene containers at 37 °C under sterile conditions [38]. The evolution of the solids surfaces were analyzed by Fourier transform infrared (FTIR) spectroscopy and scanning electron microscopy (SEM).

### 2.6 Biological assessment of the gated materials

### 2.6.1 .1 E. coli DH5α culture conditions

For viability studies, bacteria *Escherichia coli (E. coli)*, cell culture DH5α was used. Bacteria cells were maintained in glycerol 15% at -80 °C. For the assays, cells were grown for 24 hours at 37 °C and under constant stirring with 5 mL of LB medium. Cells from 1 mL culture were collected by centrifugation for 30 seconds at 13000 rpm and resuspended in 1 mL of milliQ water at pH 7.6. Then a dilution of $2 \cdot 10^4$ cells·mL$^{-1}$ was prepared, in order to achieve a final concentration of $10^4$ cells·mL$^{-1}$. The same procedure was carried out for both kinetic and viability assays.

### 2.6.2 .2 Clonogenic cell viability assay with S5-Levo, levofloxacin and ε-poly-L-lysine

For the clonogenic cell viability assay with **S5-Levo**, different suspensions of **S5-Levo** containing bacteria (final concentration $10^4$ cells·mL$^{-1}$) were prepared, achieving the final solid concentrations of 800, 500, 300, 200, 160, 120, 80, 60, 40, 20, 10, 5, 1 and 0 µg solid·mL$^{-1}$. The samples were stirred at 180 rpm (37 °C) during 10 minutes. Then, the suspensions were suitably diluted with milliQ water (pH 7.6) in order to obtain a cell growth easy to quantify. Finally, 100 µL of the new dilutions were seeded in LB plates (3% agar) and incubated at 37°C for 24 hours. Then, Colony Formation Units (CFU) were quantified.

To test the cytotoxicity of the non-loaded ε-poly-L-lysine capped solid (**S5**), the same experiment was carried out with final solid concentrations of 600, 300, 200, 100, 50, 30, 20, 5 and 0 µg solid·mL$^{-1}$.

To determine the cytotoxicity of free levofloxacin and free ε-poly-L-lysine, final concentrations of 2000, 1000, 800, 600, 400, 300, 250, 200, 150, 100, 60, 30, 10 and 0 ng·mL$^{-1}$ (levofloxacin) and 600, 400, 200, 100, 80, 50, 40, 30, 10 and 0 ng·mL$^{-1}$ (ε-poly-L-lysine) were achieved. In order to test the specificity of the gates, solid **S3-Levo** (capped with ATP) was submitted to the same procedure.

### 2.6.3 .3 Human Osteosarcoma (HOS) cells culture test

HOS cells were seeded on well culture plates (CULTEK), at a density of 40000 cell per mL in Dulbecco's Modified Eagle's Medium with 10% fetal bovine serum, 1 mM L-glutamine, penicillin (200 mg·mL$^{-1}$), and streptomycin (200 mg·mL$^{-1}$), under a $CO_2$ (5%) atmosphere at 37 °C, 24 hours to reach the confluence in each cell plate. Thereafter, **S3-Dox** solid was added on the seed cells at half confluence at 200μg·mL$^{-1}$ concentration. In order to study the stimuli-responsive behavior of the ATP gate, ALP was added to the culture wells to simulate a scenario of ALP activity excess.

### 2.6.4 .4 Cell proliferation test

Cell proliferation in contact with **S3** and **S3-Dox** was determined by the MTT method. Samples were incubated for 4 hours at 37 °C and 5% $CO_2$ under dark conditions. Then, the medium was removed and 0.5 mL of isopropanol–HCl solution were added. Finally, the absorbance was measured at 460 nm.

### 2.7 Statistics

Statistics Data are expressed as means-standard deviations of experiments. Statistical analysis was performed using the Statistical Package for the Social Sciences (SPSS) version 22 software (IBM). Statistical comparisons were made by analysis of variance (ANOVA). Subsequently, post hoc analyses were carried out to correct for multiple comparisons. In all of the statistical evaluations, P< 0.05, P<0.01 were considered as statistically significant.

## 3. Results and discussion

### 3.1 Characterization results

The synthesis strategies developed in this work are aimed to design multifunctional bioceramics with two main features. The first one is the stimulation of bone tissue regeneration to restore the skeletal integrity in bone defects. In this sense, the material should exhibit high bioactive behavior, that is, the capability to osteointegrate with the host tissue through the formation of an apatite-like phase on its surfaces. The second function is to deliver drugs on demand to treat the causal pathology of the bone defect, for instance bone infection or extirpated bone tumors. In order to reach these goals we have designed two different on demand drug delivery systems, depicted in **Scheme 1**. The first one has been designed on the basis that ATP molecules acted as capping agents. For this purpose ATP formed a covalent bond with the triamine attached to the external surface of the mesoporous material, capping the pores and inhibiting cargo release (solid **S3**). Conveniently, hydrolysis of ATP molecules induced by the over-activity of external ALP would uncap the mesopores, allowing cargo delivery in a controlled and selective way. The second system was designed by functionalizing the external surface of the MBG with 3-(triethoxysilyl)propyl isocyanate (**S4**). The resulting solid was treated with ε-poly-L-lysine, to form urea bonds with the isocyanate groups attached to the external surface to yield the final material **S5**. ε-poly-L-lysine would cover the surface of the loaded solid, capping the entrance to the pores and inhibiting payload release. Otherwise, the presence of bacterial proteases would induce hydrolysis of the amide bond in ε-poly-L-lysine, allowing cargo release. A diagram of all the prepared solids is shown in **Scheme 2**.

FTIR spectroscopy was carried out after each functionalization stage, confirming the presence of the corresponding functional groups for solids **S1, S2, S3, S4** and **S5** (see **Figure S1** in Supporting Information).

The MBG functionalization was followed by solid state NMR to confirm the successful incorporation of the different components of the gates. **Figure 1** shows solid-state $^{29}$Si single pulse (left) and cross-polarization (right) MAS NMR spectra for solids **S1, S3** and **S5.** These experiments were used to evaluate the presence of the organosilanes linkers bonded to the MBG surface. In solid **S1**, single pulse spectrum shows three signals corresponding to $Q^n$ inorganic silica environment. The observed chemical shifts, listed in **Table S1** (see Supporting Information) showed signals corresponding with $Q^4$ between -109 and -112 ppm, $Q^3$ between -100 and -103 ppm and $Q^2$ between -91 and -94 ppm. In solid **S3**, the resonances at around -58 and -67 ppm represent silicon atoms in positions $(\equiv SiO)_2Si(OH)R$ and $(\equiv SiO)_3SiR$, denoted as $T^2$ and $T^3$, respectively, evidencing the covalent functionalization with 3-[2-(2-aminoethylamino)ethylamino]propyl-trimethoxysilane. These signals are emphasized in the CP spectra pointing out that the liker is mainly located at the MBG surface. In the case of solid **S5**, the functionalization with 3-(triethoxysilyl) propylisocyanate seems to be lower, as SP spectra does not shown $T^n$ signals and only CP spectra evidenced the presence of this organosilane at the surface of **S5** [42].

Solid-state $^{31}P$ single pulse MAS NMR spectra for sample **S3** were used to evaluate the local environment of P (**Figure 2**). The band at 1.8 ppm is associated with inorganic $PO_4$ tetrahedrons within silica network. Bands at -7.2, -12.0 and -20.6 ppm are assigned to α, β and γ P atoms of the ATP molecules bounded. The higher intensity of the signal at -7.2 ppm is attributed to the convolution of the $P_γ$ of ATP with the $q^1$ signal corresponding to environmental P-O-Si bonds [43,44].

Solid-state $^1H \rightarrow ^{13}C$-CPMAS for solids **S3** and **S5** are shown in **Figure 3**. The spectra provide clear evidence that both solids were functionalized with their corresponding gates. Solid **S3** presents characteristic bands of amino-carbon bond associated with 3-[2-(2-aminoethylamino)ethylamino]propyl-trimethoxysilane between 10 and 60 ppm. Characteristic bands of ATP appear from 60 ppm and correspond with the carbons of ribose and the aromatic heterocycle. **S5** shows bands associated with 3-(triethoxysilyl)propylisocyanate and isocyanate carbon no reacted of **S4**. The spectra also show characteristic band of ε-poly-L-lysine as amide carbons at 170 ppm.[45] The assignations of carbons are summarized in **Table S2** (Supporting Information).

The mesoporous structure of the different solids was confirmed by TEM (**Figure 4**). In the case of solid **S1**, the characteristic channels of a mesoporous ordering matrix were observed as alternate black and white stripes, evidencing the 2D hexagonal ordered mesoporous structure (Figure 4A and 4B). The TEM images of solids **S3** and **S5** (figures 4C and 4D, respectively) also show ordered mesoporous arrangement, indicating that the loading and capping processes undergone by the MBG did not significantly affect the mesoporous structure.

The $N_2$ adsorption-desorption isotherms of **S1** show a type IV curve, characteristic of mesoporous materials (**Figure 5**). The curve shows a H1 hysteresis loop indicating that mesopores have an opened at both ends cylinder morphology. Solids **S3-Ru** and **S5-Ru** exhibit isotherms corresponding to mesoporous materials with lower surface area and pore volume, as could be expected after loading with $[Ru(bpy)_3]^{2+}$ and subsequent capping with the ATP (solid **S3-Ru**) or ε-poly-L-lysine (solid **S5-Ru**) based gates (see **Table 2**). Solid **S3-Ru** shows a strong decrease of textural values respect to pristine solid **S1**. In fact, the hysteresis loop changes to type H2 pointing out that the mesopore morphology shifts from open cylinders towards ink-bottle morphology as a consequence of the highly effective capping with the ATP gate. On the other hand, the textural parameters of solid **S5-Ru** underwent a lower decrease of textural parameters. The higher organic content of solid **S3-Ru** compared to **S5-Ru** calculated from thermogravimetry would justify the different decrease of porosity observed in these solids (see **Figure S2** in Supporting Information for further details).

### 3.2 Kinetic studies

In order to investigate the gating properties of ATP and ε-poly-L-lysine gates, cargo-release studies were carried out with **S3-Ru** and **S5-Ru**. In a typical experiment, 2 mg of **S3-Ru** were suspended in water at pH 7.6 in the presence and absence of ALP. Suspensions were stirred at 400 rpm at 37 °C for 10 hours, and at given time intervals fractions of both suspensions were taken and filtered to remove the solid. Dye released to the solution was then monitored by measuring the fluorescence of $[Ru(bpy)_3]^{2+}$ at 594 nm ($\lambda_{ex}$ 454 nm). $[Ru(bpy)_3]^{2+}$ delivery profiles in both, the presence and absence of ALP enzyme, are shown in **Figure 6A**. A negligible dye release occurs when ALP is not present. Contrarily, the concentration of $[Ru(bpy)_3]^{2+}$ in the solution increased significantly in the presence of the enzyme. This behavior is consistent with a tight pore closure by ATP, which would be hydrolyzed in the presence of ALP, thus unblocking the entrance to the pores and allowing cargo release.

Similar release studies were also performed with the ε-poly-L-lysine-capped solid **S5-Ru** but using pronase enzyme as trigger (**Figure 6B**). Delivery profiles in both, the presence and absence of pronase enzyme are shown in **Figure 6B**. A poor $[Ru(bpy)_3]^{2+}$ release was observed in the absence of pronase, while a significant increase of payload delivery occurs when the enzyme is added to the medium. The differences in both release kinetics are due to the fact that two different capping systems and enzymes are being used. ε-poly-L-lysine is a long polymer with a number of amide bounds to be hydrolysed, yielding L-lysine molecules. The action of the pronase enzyme allows a fast hydrolysis of the molecular gate and a quick leakage of the dye out of the pores. However, alkaline phosphatase is expected to hydrolyze the phosphate groups of ATP. This anion, in solid S3-Ru, is coordinated with positively charged amino moieties located onto a dense network of polyamines grafted onto the outer surface of the loaded support. For this reason, alkaline phosphatase could have some steric hindrance in the hydrolysis of the phosphate groups of ATP and, as a consequence, a slower release of the dye was observed.

These experiments confirm that ε-poly-L-lysine is a suitable capping system in MBG supports, which can be hydrolyzed in the presence of pronase enzyme, thus resulting in cargo release.

### 3.3 In vitro bioactivity tests

MBGs are featured by their excellent bioactive properties. Their high surface area and porosity enhance the ionic exchange with the surrounding media, thus developing an apatite-like phase onto their surface that ensures the integration with the bone under *in vivo* conditions. Up to date, MBGs have demonstrated to be the fastest bone grafts in developing this apatite like phase. However, the bioactivity of MBGs can be seriously harmed if the textural properties are modified. The ATP and ε-poly-L-lysine based gates keep the mesopores capped until the enzymatic stimuli open them. Thus, we decided to assess the bioactive behavior of solids **S3** and **S5**. For this purpose, both solids were soaked in simulated body fluid (SBF) at 37 °C to test their capability for nucleating and growing a newly formed apatite-like phase. The formation of this apatite phase was followed through the evolution of the absorption bands at 590-610 cm$^{-1}$, which appear in the FTIR spectra when this process takes place [46,47]. In addition the surface evolution and the changes in the Ca/P ratio was followed by SEM/EDX to confirm the in vitro formation of the newly growth apatite on the materials surface. The tests were carried out in absence and presence of the corresponding enzymatic stimuli. In absence of stimuli, the SEM micrograph does not evidence significant changes at the solid S3 surface (**Figure 7A**, **7B**). However, in the presence of ALP (**Figure 7C**), i.e. when the gate is open, the MBG develops a new phase after 72 hours in SBF with a Ca/P ratio of 1.66, which is very close to that of hydroxyapatite. FTIR spectroscopy agrees with the

SEM results in the absence and presence of ALP (**Figure 7D**). In the absence of ALP, the spectra show the same absorption bands before and after soaking in SBF, *i.e* those corresponding to the vibrations of Si-O bonds and P-O bonds in amorphous environment. On the contrary, in the presence of ALP, the FTIR evidence the splitting of the signal at 590-610 cm$^{-1}$ in the FTIR spectra, pointing out that the newly phase observed by SEM is an apatite-like phase. On the other hand, solid **S5** exhibits an excellent bioactive behavior. This solid develops a thick newly formed apatite phase in both the presence and absence of enzymatic stimuli, as evidenced by SEM (**Figure 7F, 7G**) and confirmed by the doublet at 590-610 cm$^{-1}$ appeared in the FTIR spectra (**Figure 7H**).

The inhibition of the bioactive behavior after the incorporation of the ATP gate could be related with the significant decrease of textural properties undergone by the MBG. The functionalization strategy used for the preparation of **S3** was very efficient, as could be observed by NMR and porosimetry measurements. The capping of the mesopores would impede the ionic exchange with the surrounding SBF thus avoiding the subsequent nucleation and growth of the newly formed apatite phase. In order to find out the significance of this effect, we analyzed the $Ca^{2+}$-$H^+$ exchange of this solid during the first stages of the bioactive process (**Figure S3**). The $Ca^{2+}$ release from solid **S3** to the SBF is significantly decreased compared with MBG without capping. $Ca^{2+}$ release is the first reaction required to initiate the bioactive process [48,49]. The inhibition of this stage seriously hiders the rest of the reactions that lead to the formation of an apatite-like phase similar to the mineral component of the bone. Another possible explanation could be the entrapping of $Ca^{2+}$ by the phosphate groups of the ATP gate. It has been highly demonstrated that the affinity of phosphate groups for $Ca^{2+}$ can inhibit its release (and consequently the bioactivity) by forming CaP nanoclusters that inhibit the $Ca^{2+}$ dissolution. The CaP clusters formation has been widely studied within the walls of the MBGs [50–53] and we hypothesize that the $Ca^{2+}$ entrapment by phosphates could also occurs at the MBG surface due to the ATP presence. Independently of the mechanism that inhibits the bioactivity of **S3**, ALP open the ATP based gate and the MBG recovers its bioactivity.

**3.4 Biological assessment of the gated materials**

In order to test the stimuli-responsive properties of ATP-gated materials, the release of doxorubicin from **S3-Dox** solid was analyzed in the absence and presence of ALP (Figure 8). A negligible drug release occurs in the absence of ALP, whereas a marked doxorubicin release was observed when ALP is added, pointing out that the enzyme opens the ATP gates.

The antitumoral activity of **S3-Dox** was studied with HOS cell cultures. HOS viability was measured in the presence of **S3** and **S3-Dox** under absence and presence of ALP (**Figure 9**). The cells in contact with **S3** show a proliferative behavior very similar to those cultured on the polystyrene control. However, the cells cultured with **S3-Dox** clearly underwent a cytotoxic effect. In the absence of ALP, the HOS proliferation is significantly hampered after 24 hours respect to the control. This result can be due to the small amount of doxorubicin released, even without the addition of ALP. It must be taken into account that HOS cells produce ALP by themselves and could facilitate the partial opening of the ATP gates. However, 48 hours later HOS proliferate indicating that the doxorubicin released is not enough to inhibit the growth of HOS. On the contrary, in the presence of ALP, HOS cannot proliferate after 48 hours of culture showing a significant viability decrease as a consequence of the released doxorubicin. Finally, the HOS viability was determined for longer exposure times. After 144 hours of test, the HOS viability in the presence of solid **S3** remains unaltered. However, in the case of doxorubicin loaded sample (**S3-Dox**), a significant decrease of HOS viability was observed under both conditions, i.e. ALP presence and absence. This fact can be attributed to the MBG solubility that allows the partial release of the antitumoral drug. However, it must be highlighted that doxorubicin release is still higher in the presence of ALP, pointing out that the stimuli-responsive behavior is present even after the partial degradation of the MBG matrix associated to the bioactive process.

In a second scenario, the MBG solid capped with ε-poly-L-lysine was loaded with the antibiotic levofloxacin (solid **S5-Levo**) and the selective delivery of the antibiotic in the presence of bacteria was evaluated. For this study *Escherichia coli,* DH5α strain, was used as model bacteria. In a first step, the amount of cargo released from **S5-Levo** was determined in the presence and in the absence of bacteria. As shown in **Figure 10**, a limited payload delivery (less than 20% after 25 hours) was found in the absence of *E. coli*, whereas a remarkable cargo delivery was found when the bacteria were present. From these studies it was found that maximum amount of levofloxacin released from **S5-Levo** was of 0.322 ng of levofloxacin per mg of solid. Moreover, it was also observed that delivery of antibiotic levofloxacin was concomitant with a reduction of bacteria viability (*vide infra*).

The antimicrobial activity of **S5-Levo** was studied more in detail by carrying clonogenic cell-viability assays in which *E. coli* bacteria were treated with different concentrations of **S5-Levo** at pH 7.6. A negative assay with no bacteria was also carried out and used as control to quantify cell growth. In a typical experiment, bacteria ($10^4$ cells·mL$^{-1}$) were incubated for 5 minutes in the presence of **S5-Levo**, and then seeded in petri plates. Seeded plates were incubated at 37 °C for 24 hours and then colony formation units (CFU) were quantified. **Figure 11** shows %CFU versus the maximum amount of levofloxacin that can be released from **S5-Levo** (i.e. 0.322 ng of levofloxacin per mg of solid, *vide ante*). As seen in the figure, CFU values were correlated with the amount of **S5-Levo** added to the bacteria medium. **S5-Levo** showed considerable toxicity against *E. coli*, with an EC$_{50}$ value of 22.26 ng·mL$^{-1}$. Additionally, to demonstrate that the antibiotic effect found for **S5-Levo** was due to the release of levofloxacin in the presence of *E. coli*, similar experiments were carried out with a MBG support also capped with ε-poly-L-lysine but that did not contain levofloxacin (solid **S5**). This solid showed negligible toxicity against bacteria (data not shown), strongly suggesting that the delivery of antibiotic levofloxacin from **S5-Levo** was the responsible for bacterial death. Moreover, in a further experiment the toxicity of solid **S5-Levo** was compared to that of free levofloxacin and free ε-poly-L-lysine. As depicted in **Figure 11**, levofloxacin and ε-poly-L-lysine showed similar toxicity to *E. coli* with EC$_{50}$values of 151.99 ng·mL$^{-1}$ and 131.37 ng·mL$^{-1}$, respectively. This is a remarkable result that indicated that levofloxacin is seven-fold more toxic when entrapped in **S5-Levo** than when free.

The toxic effect observed for **S5-Levo** described above against *E. coli* can be understood keeping in mind that ε-poly-L-lysine molecules can be hydrolysed by proteolytic enzymes excreted by the bacteria [54] which would result in levofloxacin delivery. In order to demonstrate that protease activity of *E. coli* was the responsible of the release of levofloxacin via degradation of the ε-poly-L-lysine cap, parallel experiments were performed with solid **S3-Levo**. This is a MBG support loaded with the antibiotic levofloxacin and capped with the ATP gate-like ensemble described above. As the ATP gate cannot be degraded by proteases, it was expected that this solid would not be toxic for *E. coli*. The antimicrobial activity of solid **S3-Levo** was studied following a similar protocol to that described above for **S5-Levo**. However, in this case, no remarkable toxicity to *E. coli* was found after treating bacteria with **S3-Levo** and no levofloxacin delivery was observed.

**4. Conclusions**

In conclusion, it has been demonstrated that MBGs can be functionalized and implemented with tailored molecular gates, allowing a new application as controlled delivery device. Specifically, two different solids were prepared from a MBG selected as an inorganic support. One portion was functionalized with a triamine and capped with ATP, whereas the other portion was functionalized with isocyanates and capped with ε-poly-L-lysine. Solids following each step were correspondingly characterized, finding evidence of their correct functionalization with the tailored molecular gates. Both molecular gated mechanisms were proved to be opened, in the presence of the corresponding enzymes (ALP and pronase, respectively), while they remained closed in the absence of the stimuli. Moreover, we have demonstrated the different bioactive behavior in both gated solids. Whereas ε-poly-L-lysine-capped system allows the formation of crystalline hydroxyapatite on their surface under any scenario, the ATP-capped system requires the opening of the molecular gate to initiate the formation of the apatite phase.

Finally, the *in vitro* efficiency of both gated systems was validated. The ATP-capped system responds to the presence of high levels of ALP, opening the gates and releasing doxorubicin. Higher serum ALP levels are found in patients developing osteosarcoma and our ATP-capped system has demonstrated to inhibit the HOS cells proliferation under a similar scenario. Regarding the ε-poly-L-lysine-capped system, we demonstrated that the presence of *E.coli* bacteria was also able to hydrolyze the ε-poly-L-lysine gate, allowing levofloxacin release. Therefore, we also tested the cytotoxic effect of the solid, and demonstrated that equitoxic concentration of the levofloxacin contained in the solid was more effective against bacteria than free levofloxacin and free ε-poly-L-lysine. We also proved that the capped solid performed no cellular damage when there was no levofloxacin inside the pores, which evidences that the cytotoxicity is totally caused by the released levofloxacin.

Thus, we have envisioned here a new approach to mesoporous gated materials, which is expected to set up innovative pathways to the treatment of bone diseases.

**Acknowledgements**

Spanish Government for projects MAT2015–64139-C04–01-R, MAT-2013-43299-R, MAT2015-64831-R (MINECO/FEDER) and for project CSO2010-11384-E (Agening-MICINN). Also, Generalitat Valenciana (project PROMETEOII/2014/047) and CIBER-BBN (project SPRING) are acknowledged for their support. MVR acknowledges funding from the European Research Council (Advanced Grant VERDI; ERC-2015-AdG Proposal No. 694160). LP thanks Universitat Politècnica de València for her FPI grant. NGC is greatly indebted to Ministerio de Ciencia e Innovación for her predoctoral fellowship. The authors also wish to thank the staff of the ICTS Centro Nacional de Microscopía Electrónica of the Universidad Complutense de Madrid (Spain) for the assistance in the scanning electron microscopy.

**Figure caption**

**Scheme 1**. Schematic representation of the ATP (A) and ε-poly-L-lysine (B) molecular gates.

**Scheme 2.** Flow diagram showing the preparation and name of the solids used in this paper.

**Figure 1.** Solid-state $^{29}$Si single-pulse (left) and cross-polarization (right) MASNMR spectra of S1, S3 and S5 solids. The areas for the $Q^n$ units were calculated by Gaussian line-shape deconvolutions (their relative populations are expressed as percentages).

**Figure 2**. Solid-state $^{31}$P single-pulse MASNMR spectra (with their respective phosphorus environments shown at the top) of solid **S3**.

**Figure 3**. Solid-state $^{13}$C single-pulse MASNMR spectra (with their respective carbon environments shown at the top) of solids **S3** and **S5**. **\*** Signal assigned to unreacted isocyanate groups.

**Figure 4.** Representative TEM images of solid **S1** acquired along [1 0] direction (A), solid **S1** acquired along the [0 1] direction (B), solid **S3** (C) and solid **S5** (D). The FT diagrams for **S1** solid are included (A and B), evidencing the 2D hexagonal ordering of this sample.

**Figure 5**. N$_2$ adsorption-desorption isotherms for solids **S1**, **S3-Ru** and **S5-Ru**.

**Figure 6.** Dye release studies carried out at 37 ºC: A) MBG capped with ATP (**S3-Ru**), B) MBG capped with ε-poly-L-lysine (**S5-Ru**).

**Figure 7.** Evolution of materials surfaces during the in vitro bioactivity test. A) SEM micrograph of solid **S3** before soaking; B) **S3** soaked in SBF for 72 hours in absence of ALP; C) **S3** soaked in SBF for 72 hours in presence of ALP (inset: magnification x 35000) , D) FTIR spectra of solid **S3** before and after being soaked in SBF for 72 hours

in the presence and absence of ALP. E) SEM micrograph of solid **S5** before soaking., F) **S5** soaked in SBF for 72 hours in the absence of pronase, G) **S5** soaked in SBF for 72 hours in presence of pronase (inset: magnification x 35000); H) FTIR spectra of solid **S5** before and after being soaked in SBF for 72 hours in the presence and absence of pronase.

**Figure 8**. Doxorubicin release studies carried out at 37 °C with **S3-Dox** in the presence and in the absence of ALP.

**Figure 9.** HOS viability in contact with **S3** and **S3-Dox** solids in the absence and presence of ALP after 6, 24, 48 and 144 hours. Significant differences from 48 and 144 hours: Statistical significance: * p<0.05, **p<0.01

**Figure 10.** Levofloxacin release studies carried out at 37 °C with **S5-Levo** in the presence and in the absence of *E. coli*.

**Figure 11.** Free levofloxacin (solid circles), free ε-poly-L-lysine (open circles), equitoxic concentration of levofloxacin in **S5-Levo** (solid squares) and equitoxic concentration of levofloxacin in **S3-Levo** (solid triangles) versus % CFU.

**Table 1.** Summary of the synthesized solids.

| Name | Support | Cargo | External functionalization | Cap |
|---|---|---|---|---|
| **S1** | MBG[a] | -- | -- | -- |
| **S2** | MBG | -- | N3[b] | -- |
| **S3** | MBG | -- | N3 | ATP |
| **S4** | MBG | -- | NCO[c] | -- |
| **S5** | MBG | -- | NCO | ε-poly-L-lysine |
| **S3-Ru** | MBG | $[Ru(bpy)_3]^{2+}$ | N3 | ATP |
| **S3-Levo** | MBG | Levofloxacin | N3 | ATP |
| **S3-Dox** | MBG | Doxorubicin | N3 | ATP |
| **S5-Ru** | MBG | $[Ru(bpy)_3]^{2+}$ | NCO | ε-poly-L-lysine |
| **S5-Levo** | MBG | Levofloxacin | NCO | ε-poly-L-lysine |

[a] MGB: Mesoporous bioactive glass
[b] N3: with 3-[2-(2-aminoethylamino)ethylamino]propyl-trimethoxysilane
[c] NCO: 3-(triethoxysilyl)propylisocyanate

**Table 2.** Textural properties of solids **S1**, **S3-Ru** and **S5-Ru.**

| Sample | $S_{BET}$ (m$^2$g$^{-1}$) | Pore volume (cm$^3$g$^{-1}$) | Pore size (nm) |
|---|---|---|---|
| **S1** | 307.10 | 0.318 | 4.2 |
| **S3-Ru** | 90.35 | 0.100 | 3.7 |
| **S5-Ru** | 268.43 | 0.263 | 3.6 |

Scheme 1
Click here to do.....oload high resolution image

A) 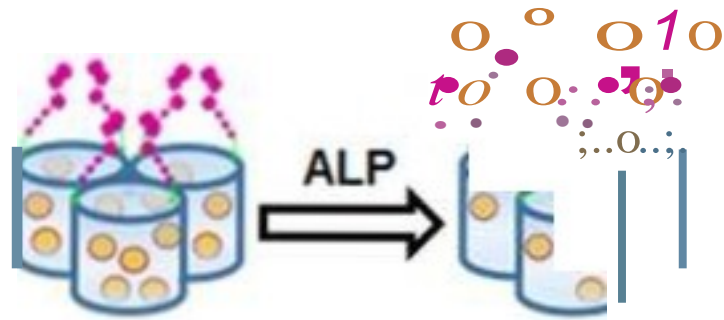

B) 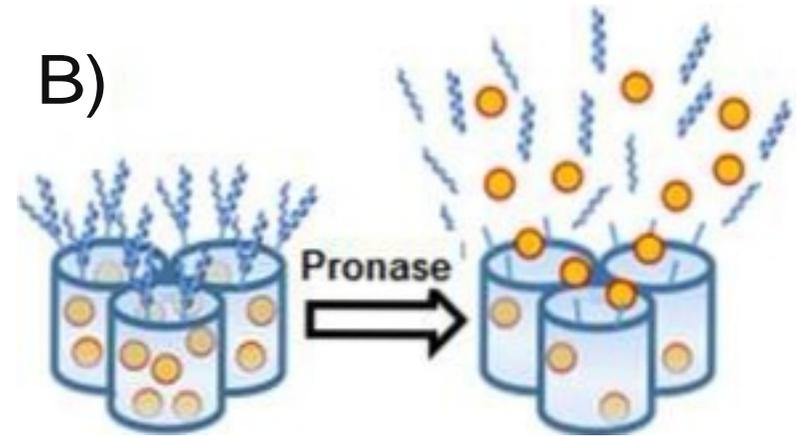

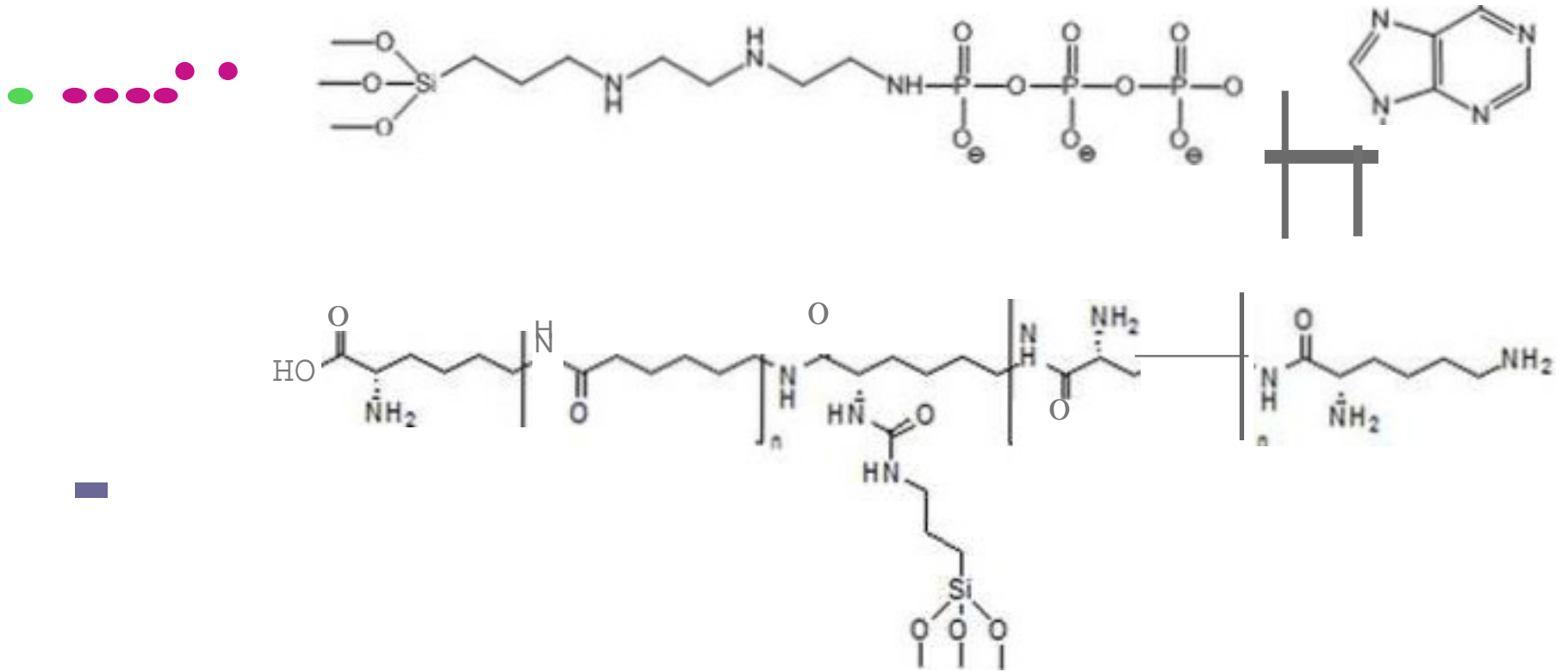

Scheme 2

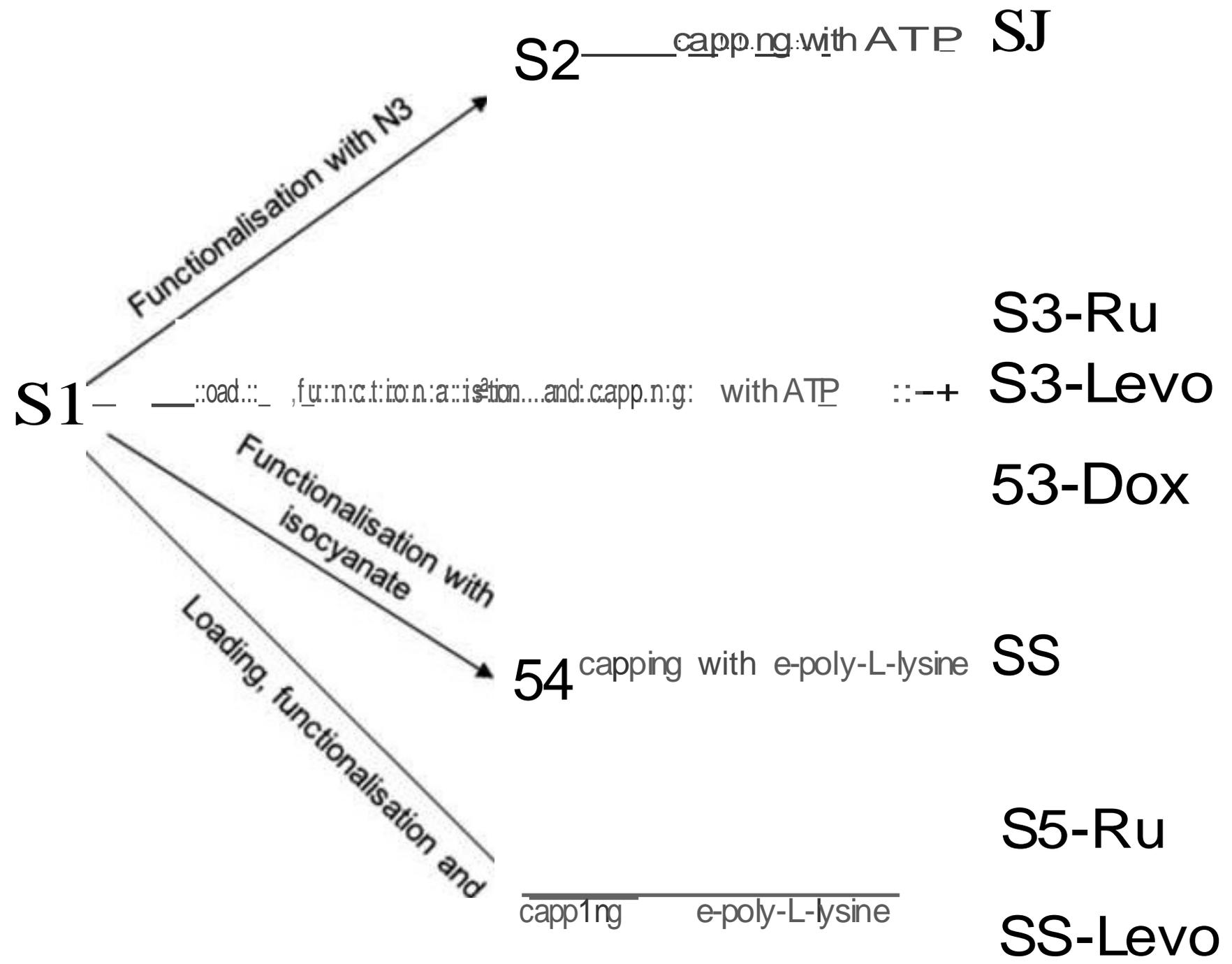

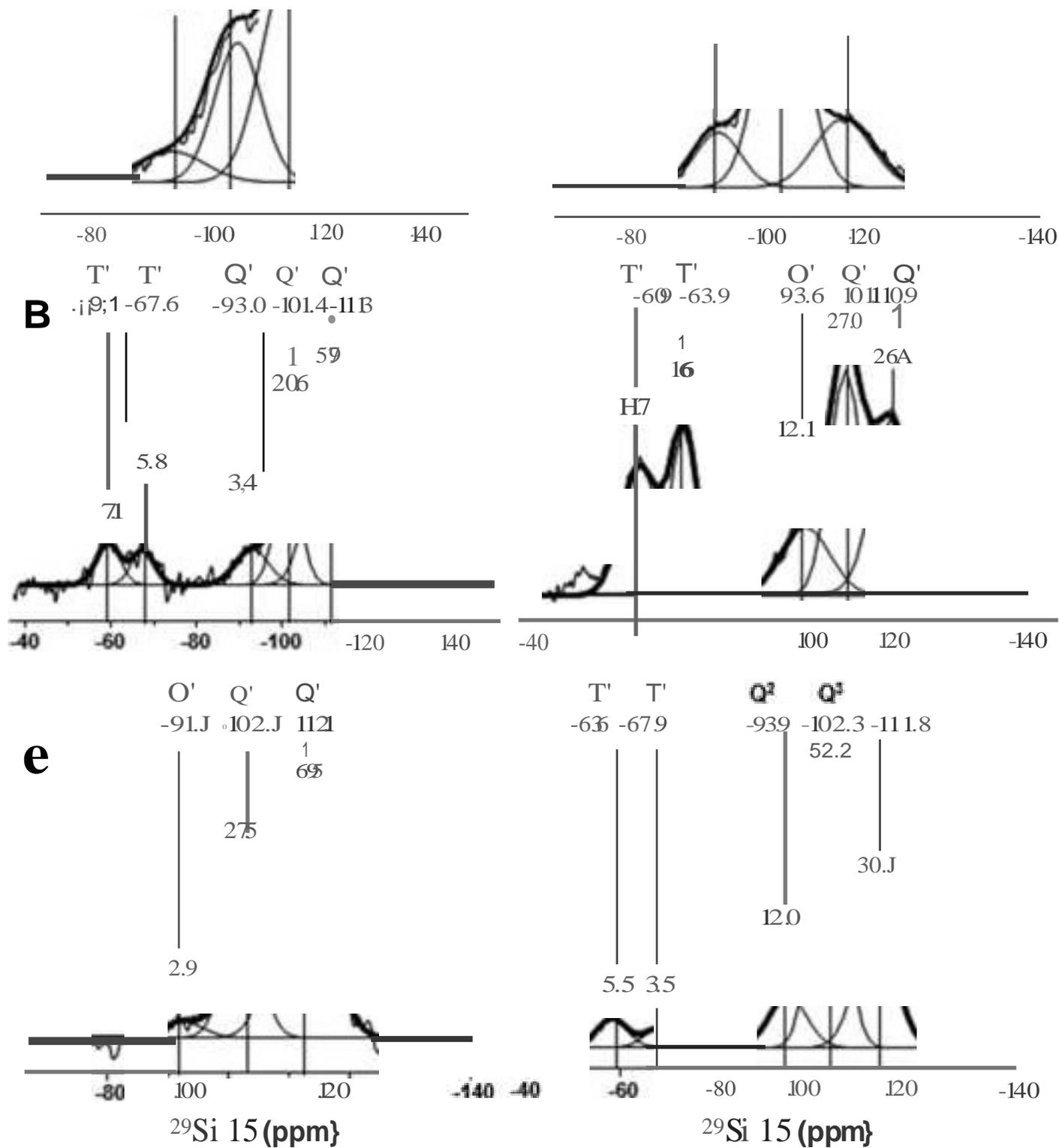

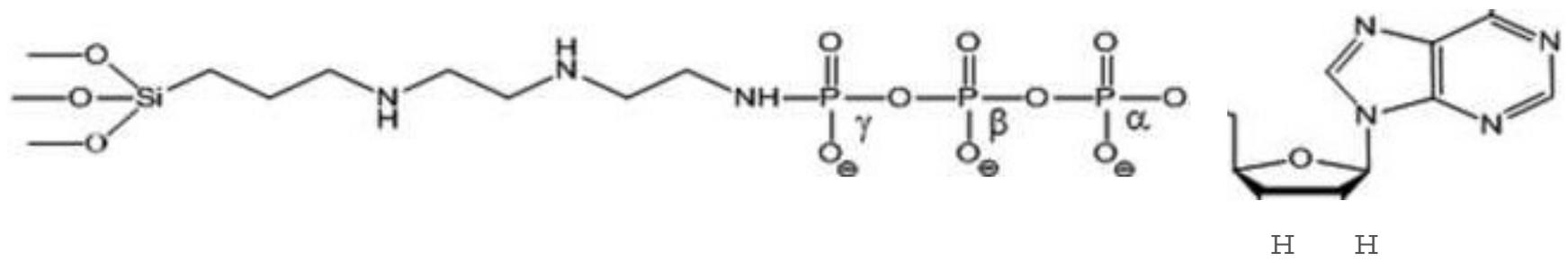

**53**

Single pulse

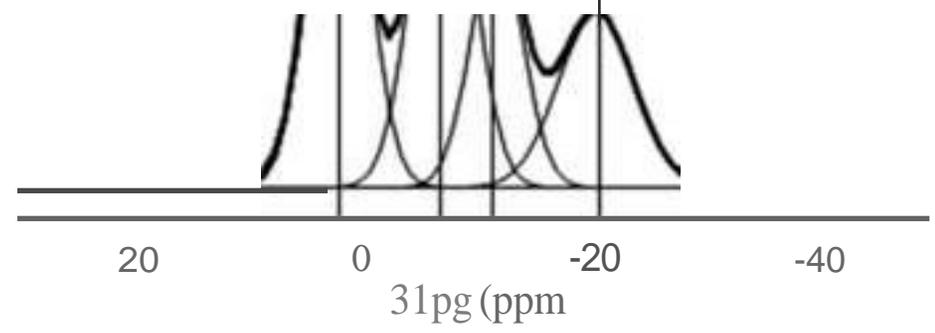

1H-+31p CP

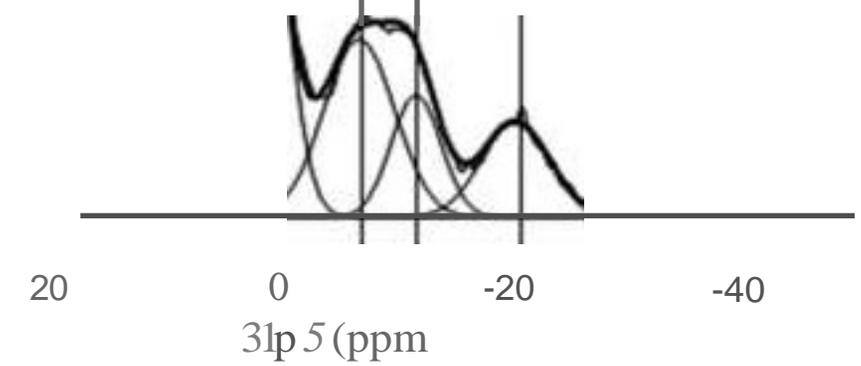

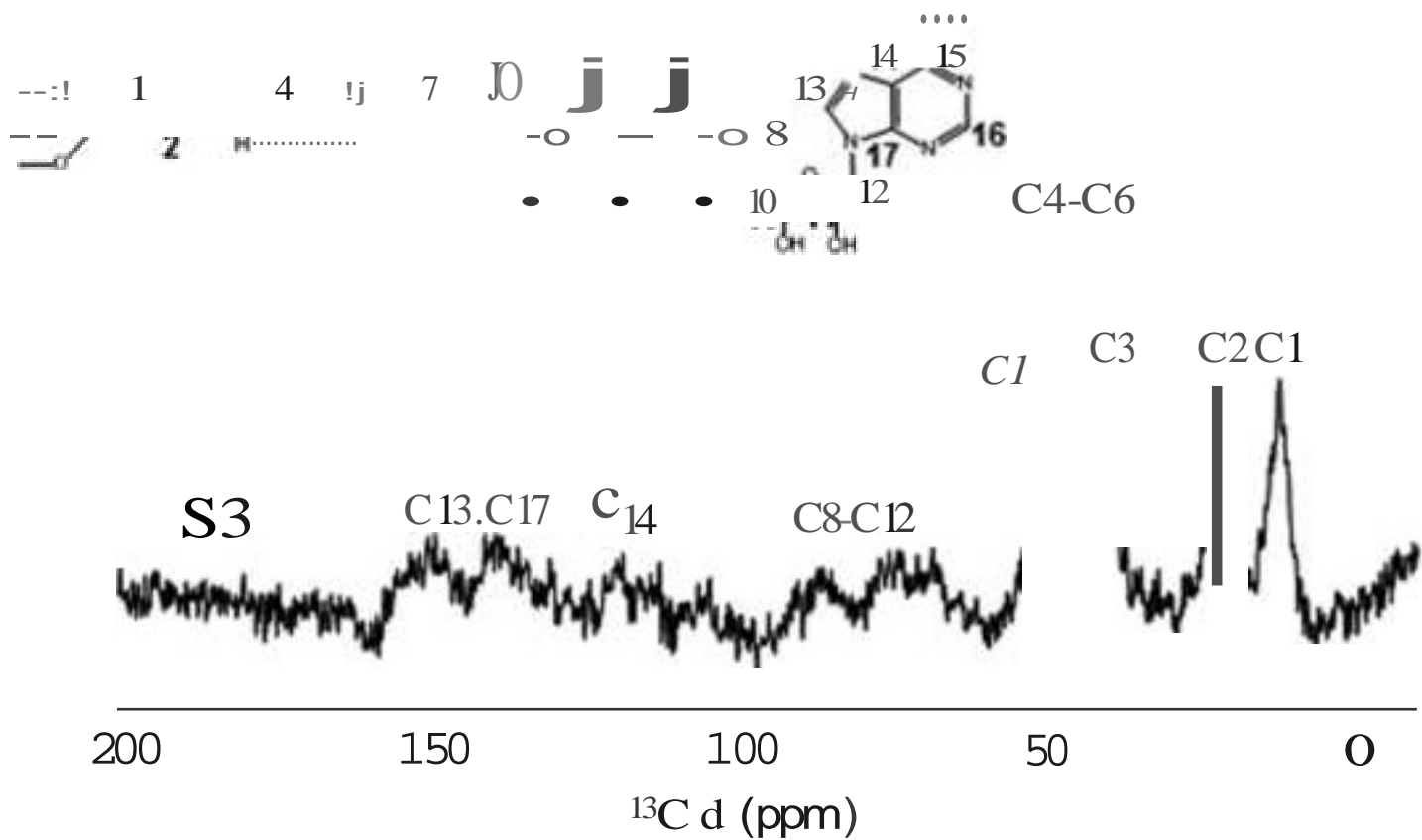
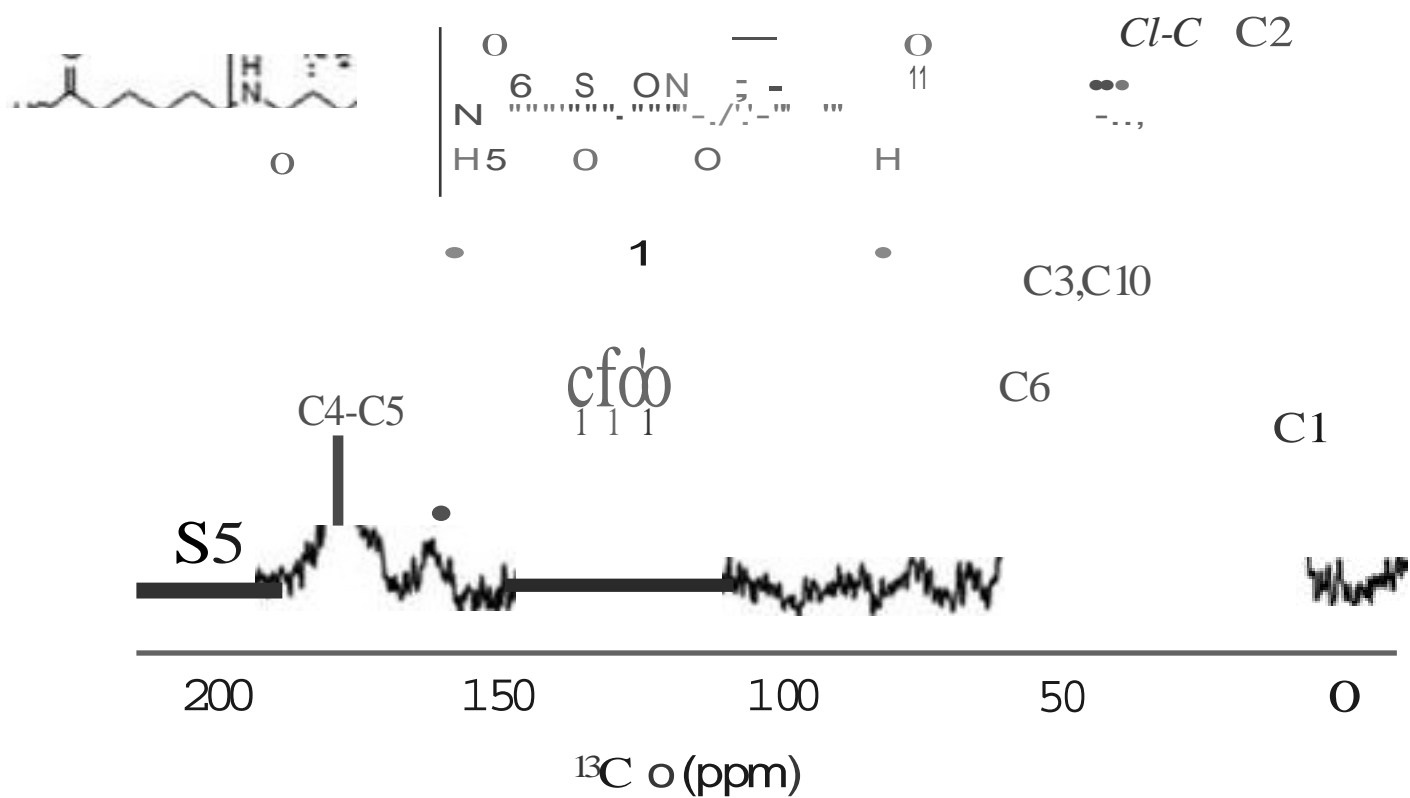

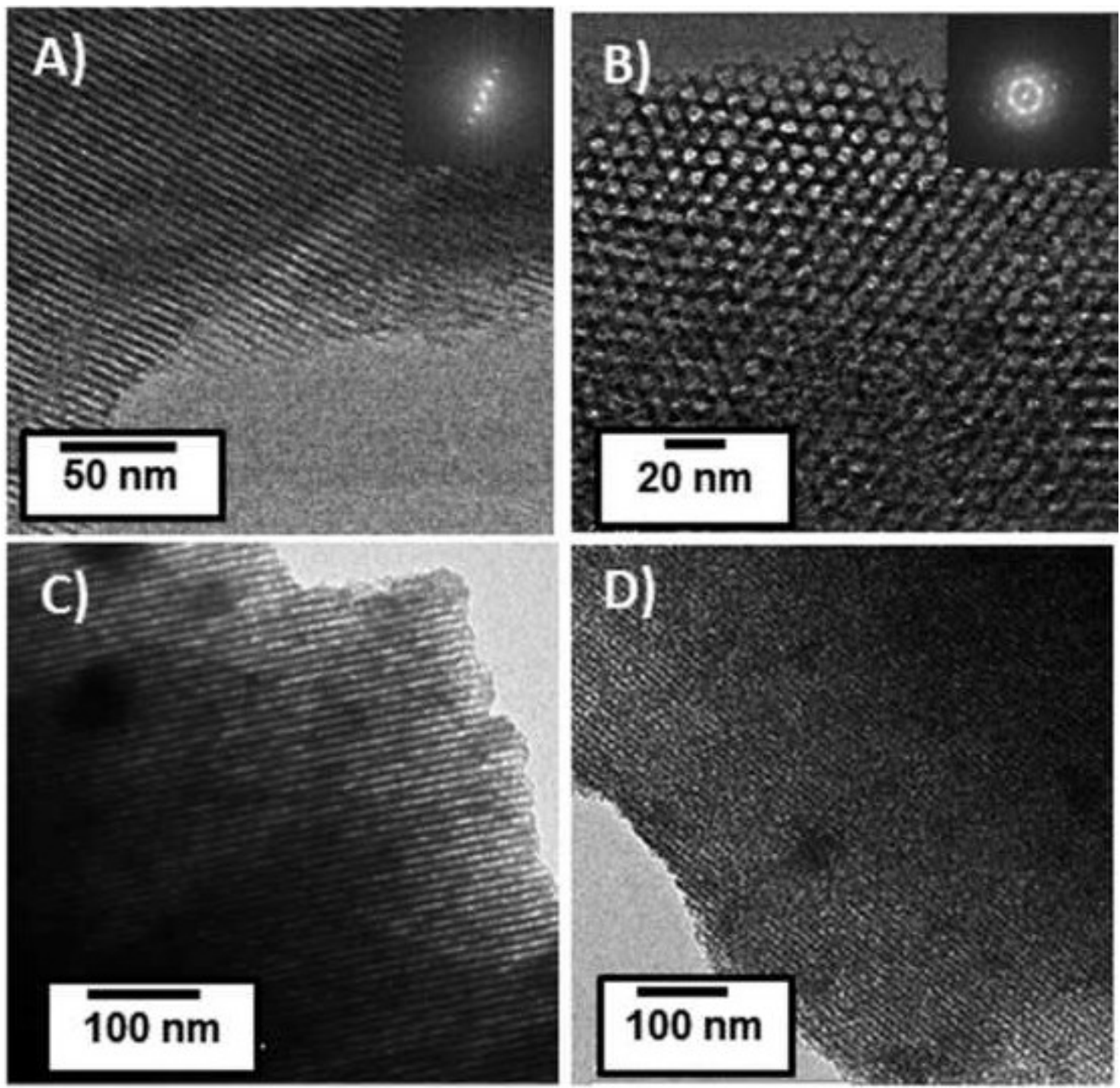

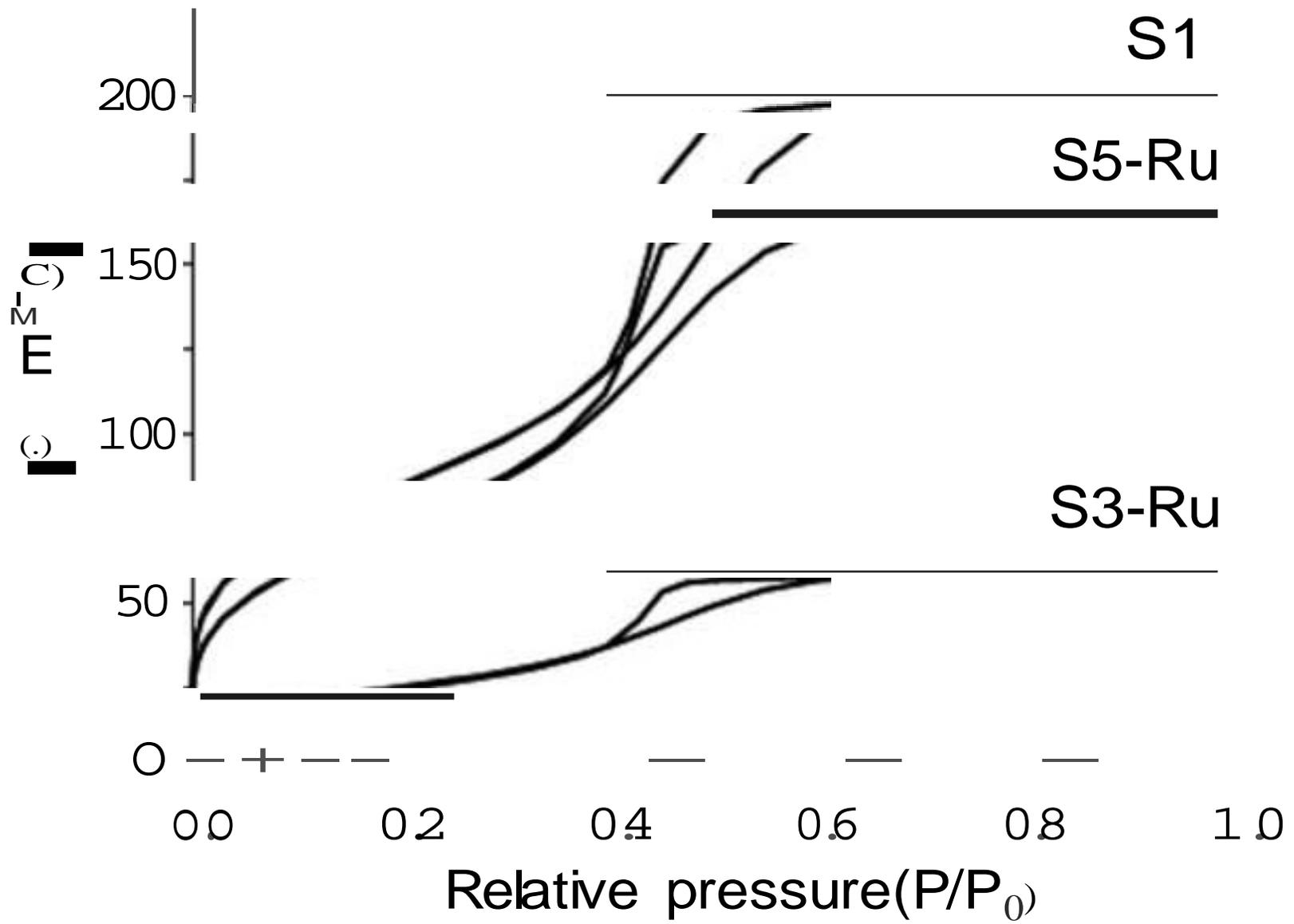

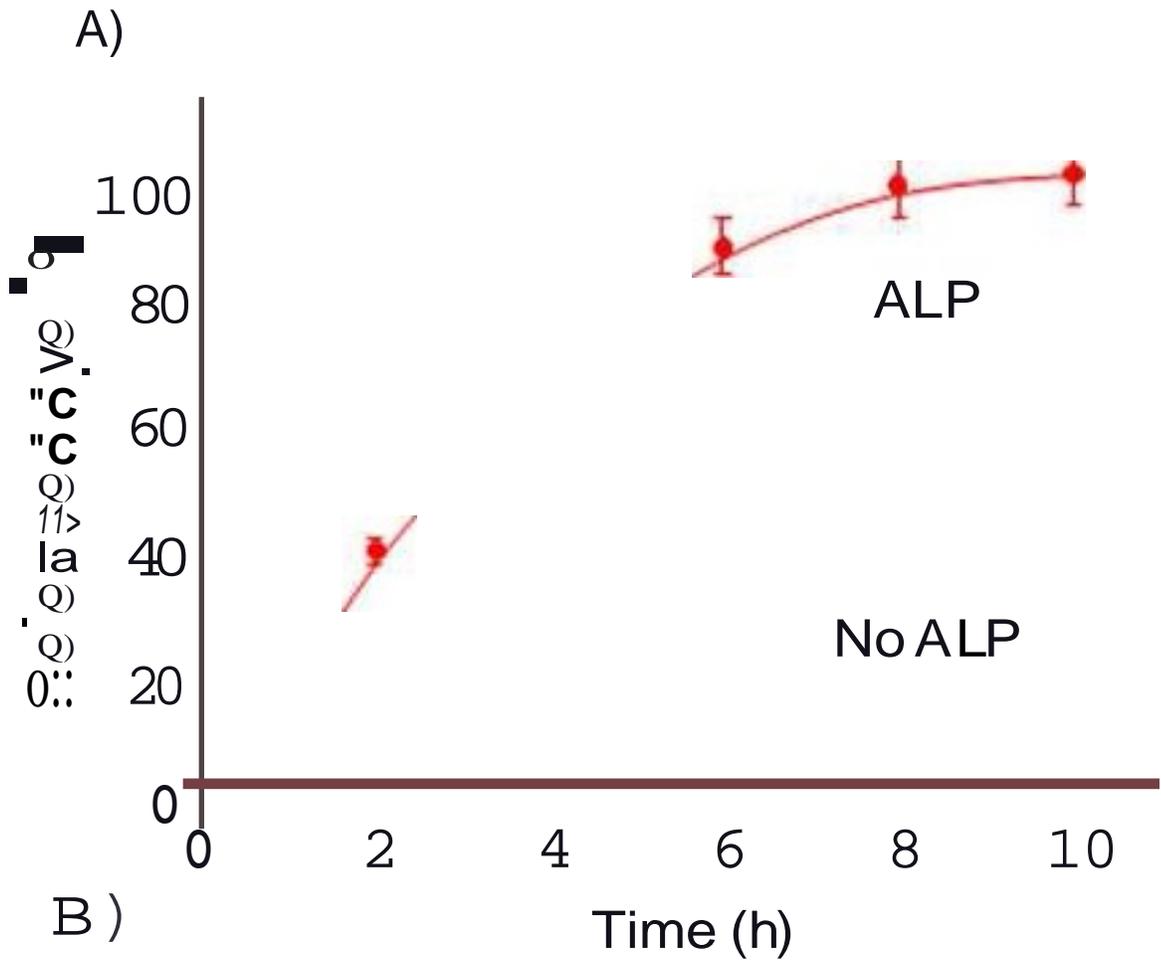

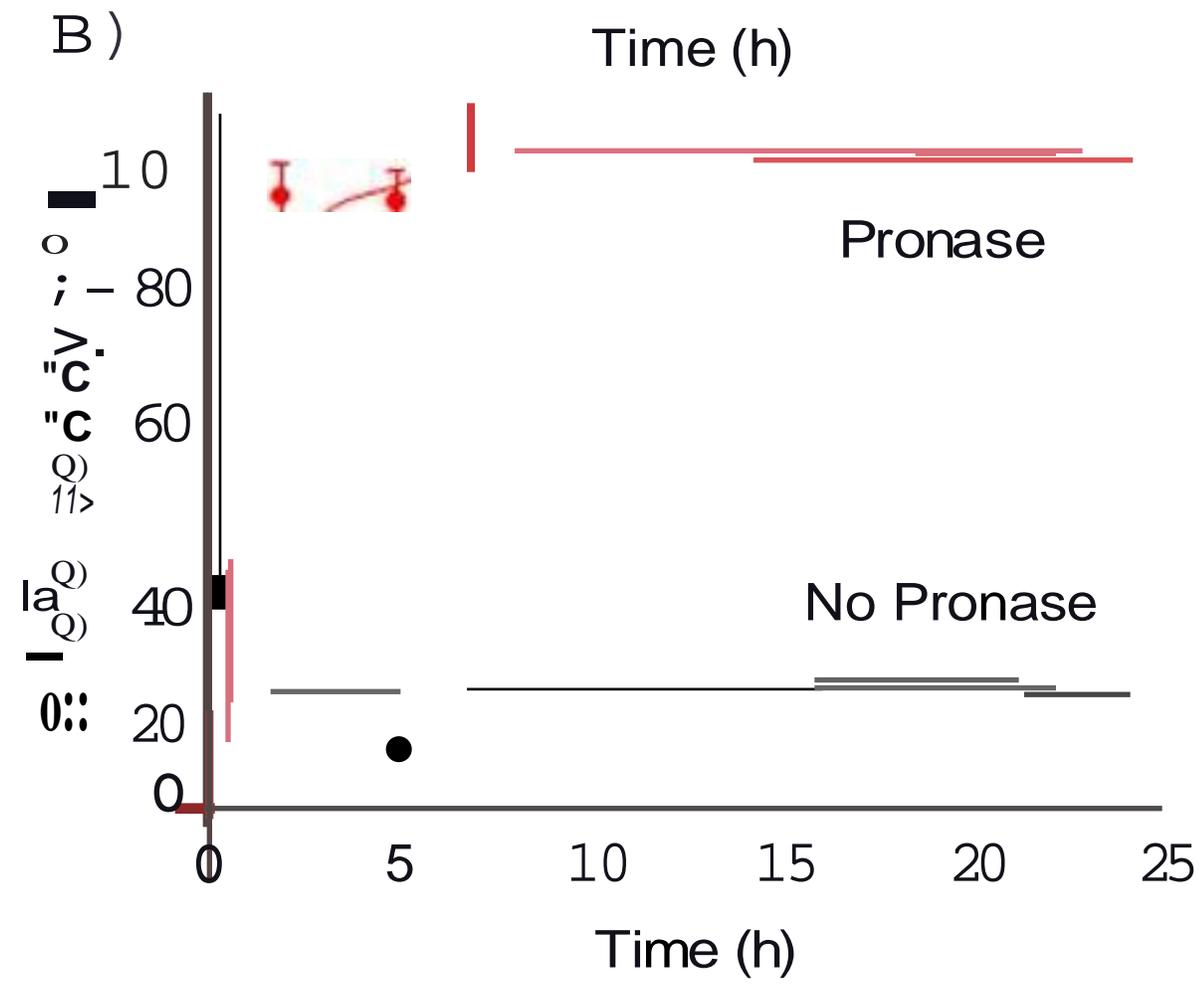

**Figure 7**

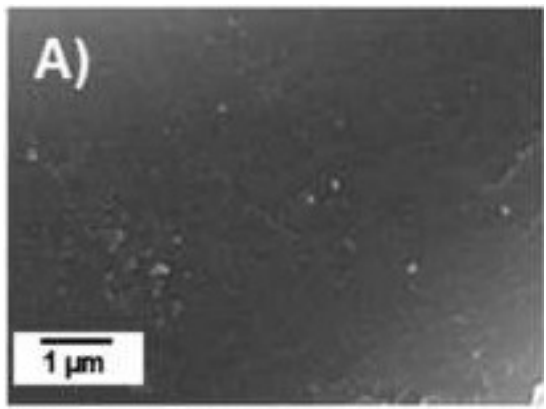
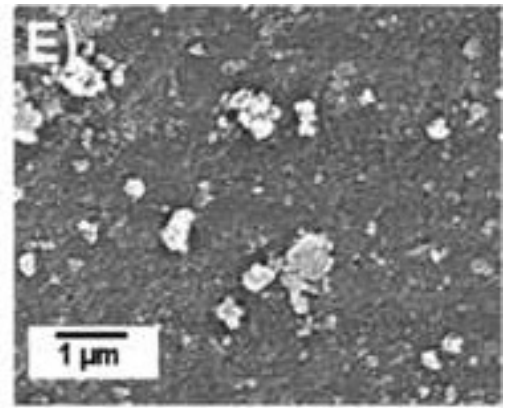
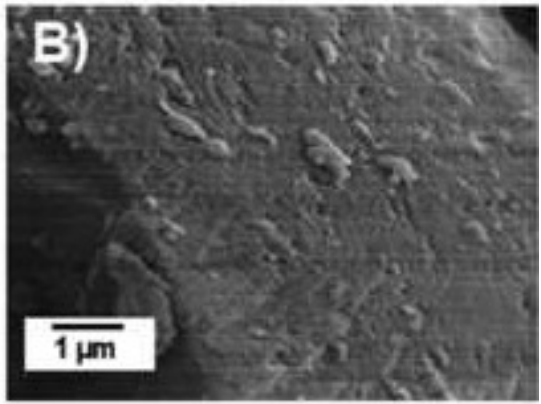
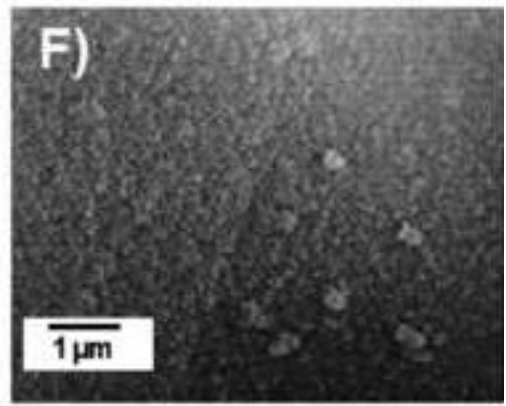
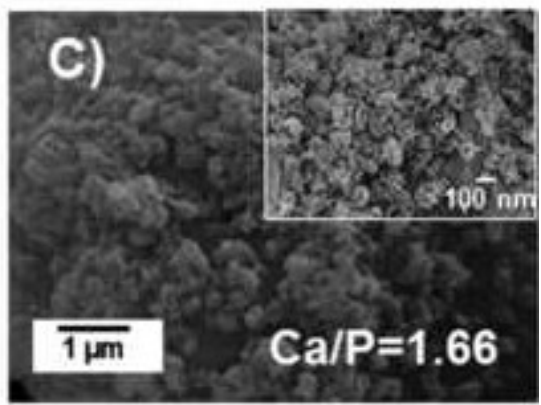
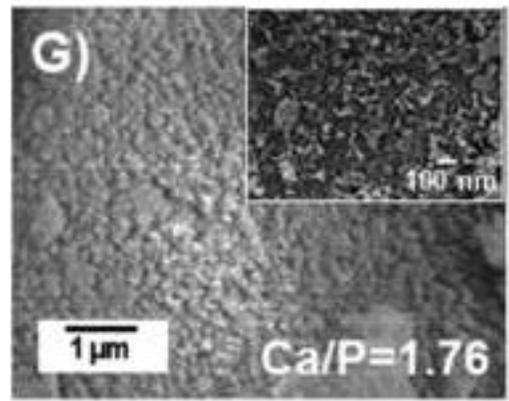
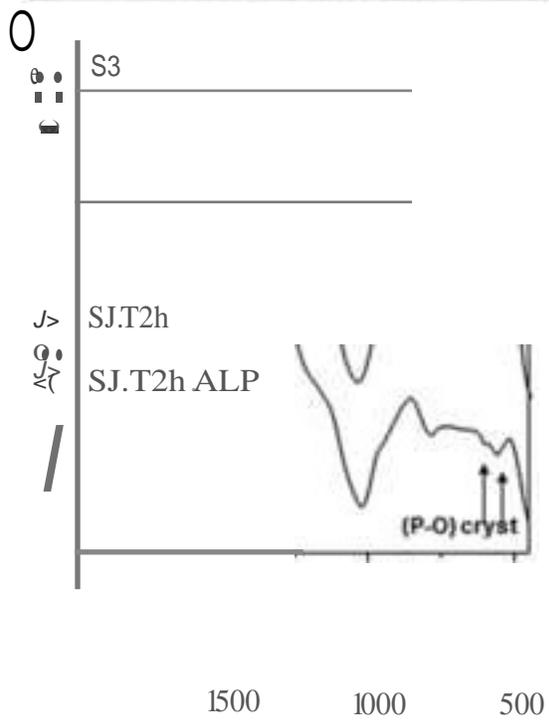
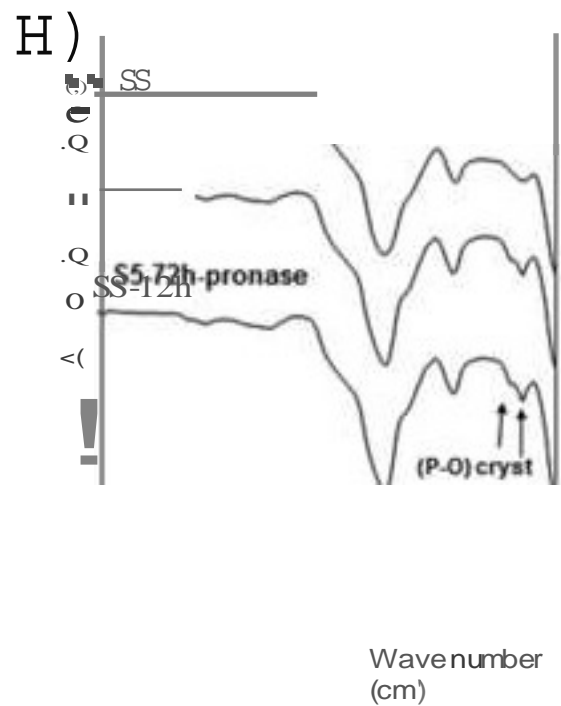

1500  1000  500
Wavenumber (cm⁻¹)

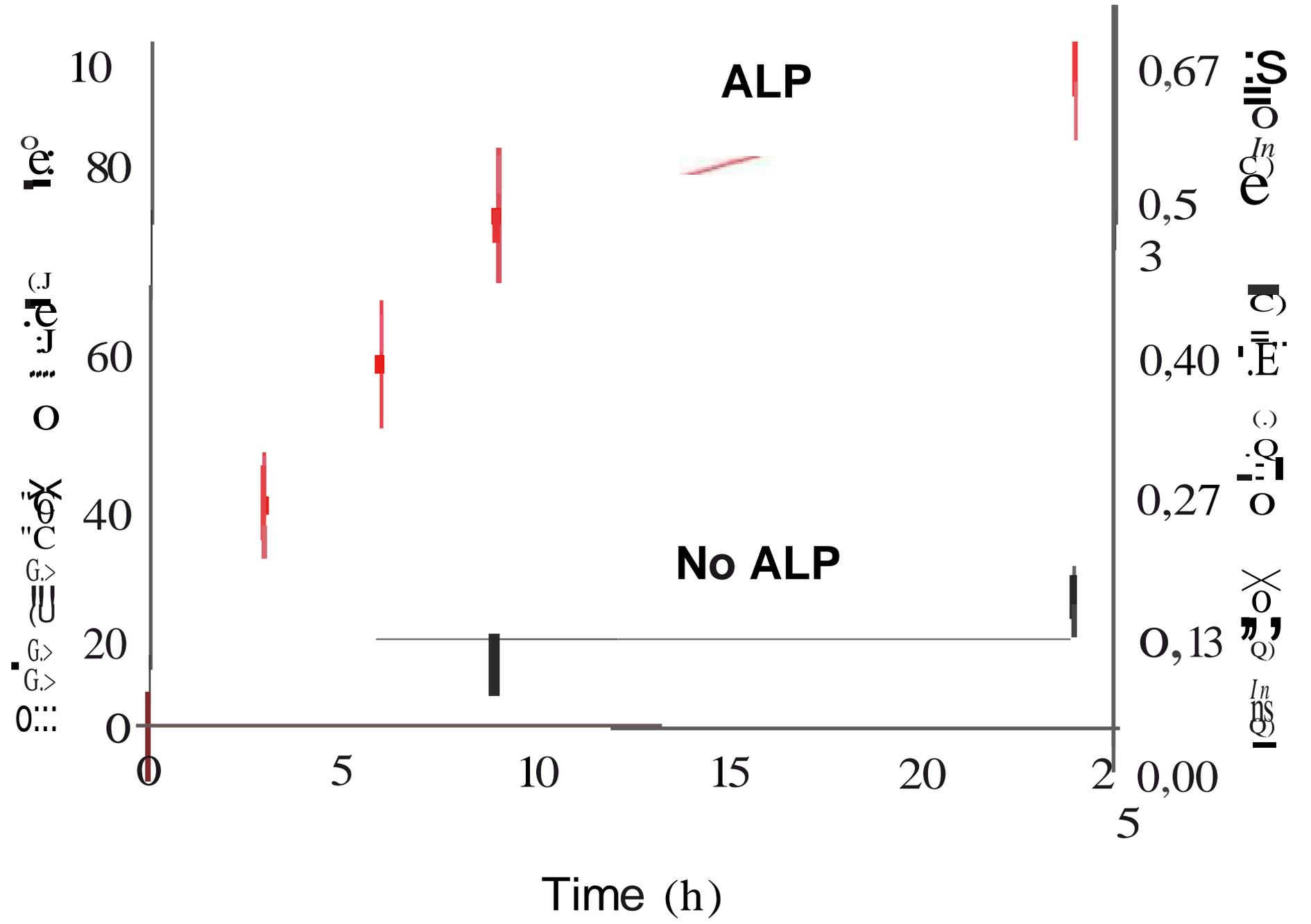

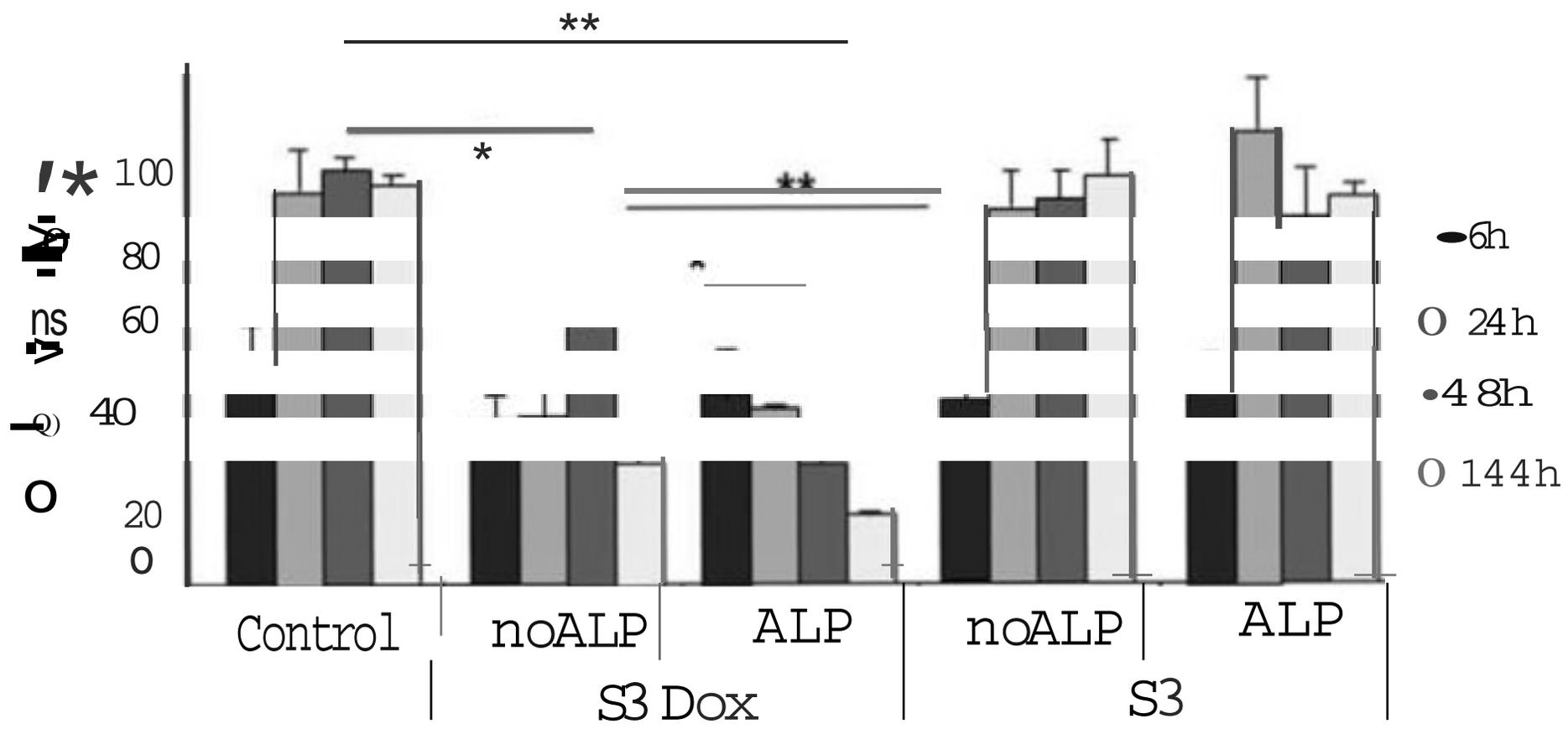

**Figure 10**
[Click here to download high resolution image]

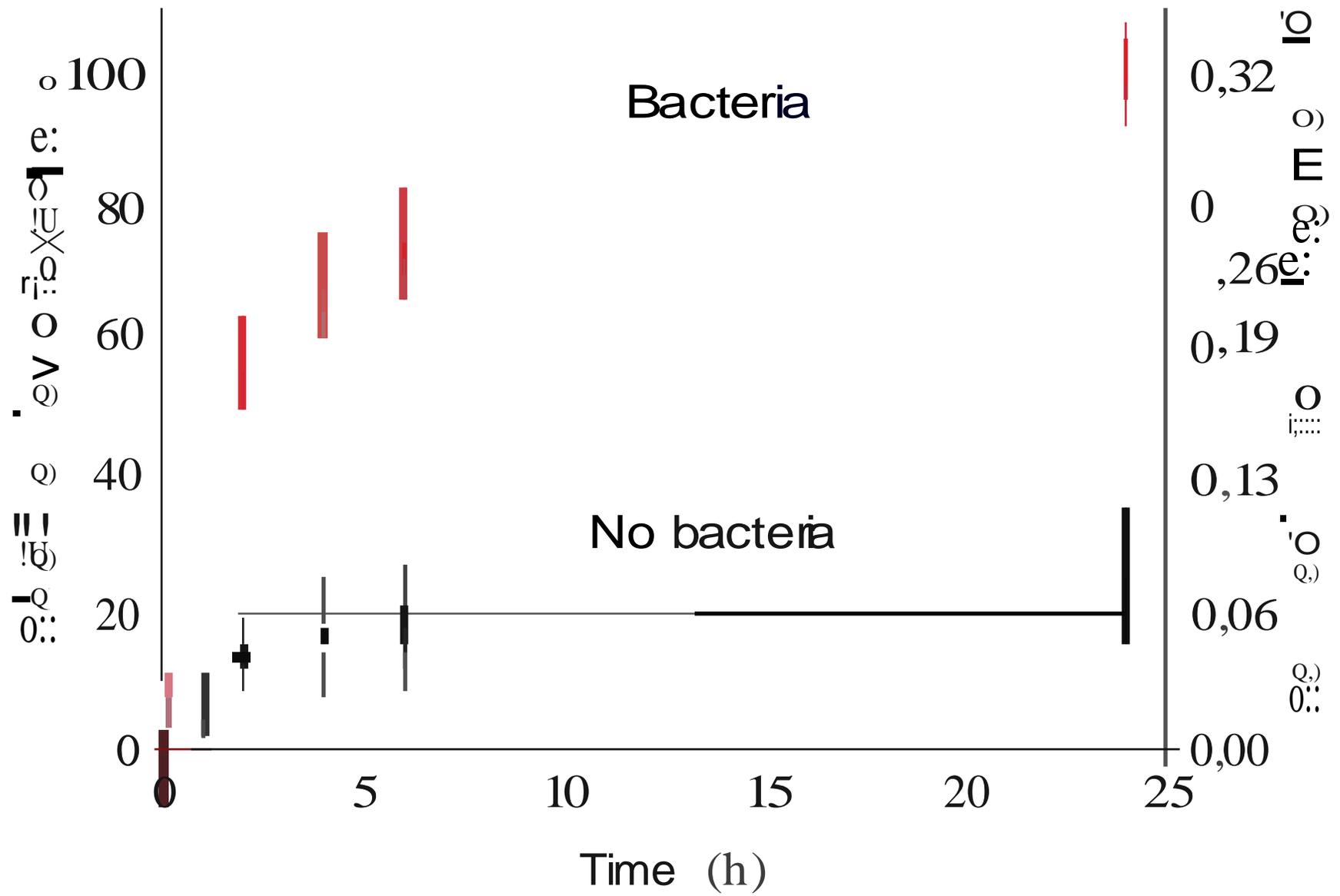

Figure 11
Click here to download high resolution image

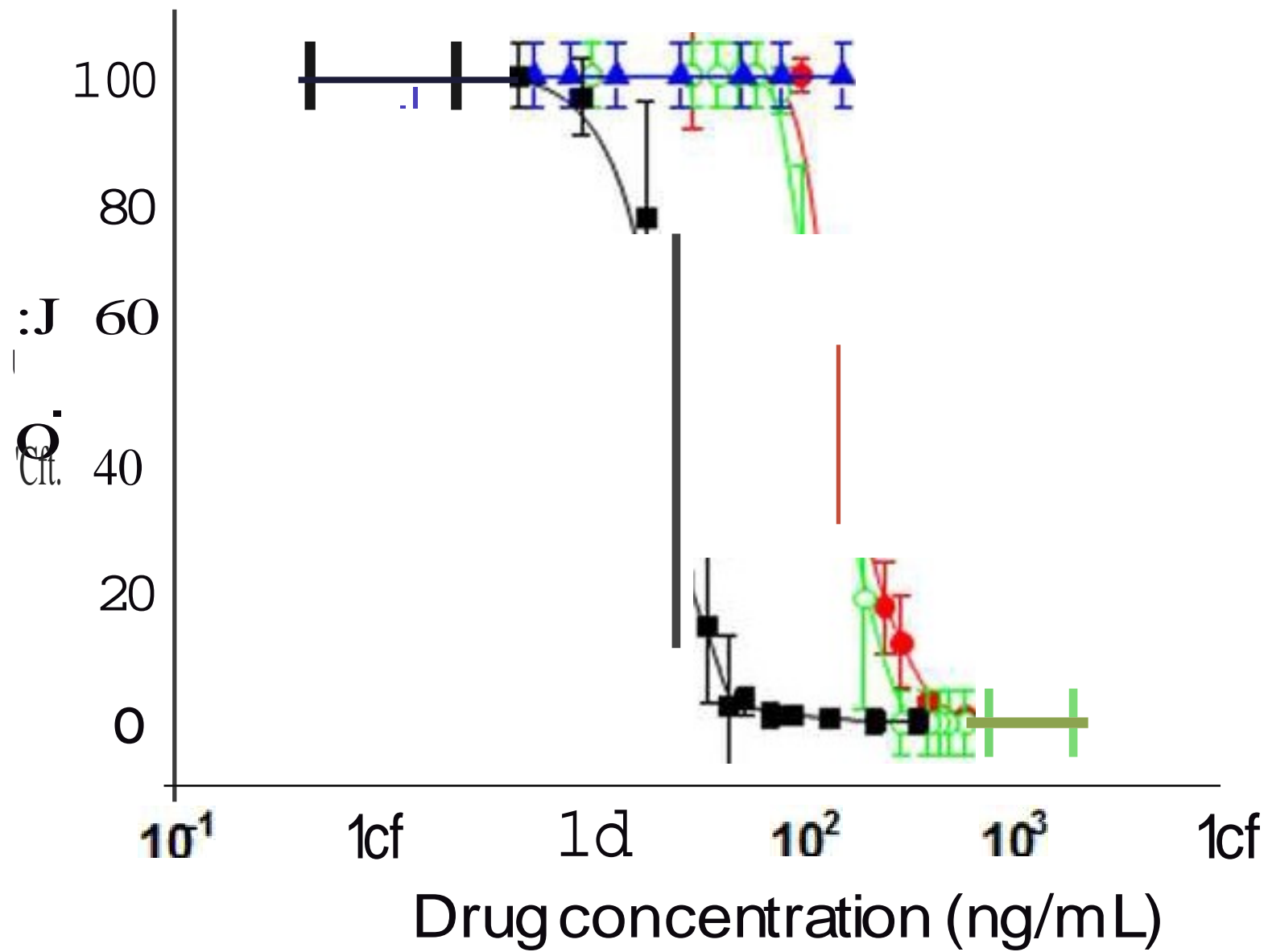